\documentclass[a4paper,11pt]{article}
\pdfoutput=0 

\usepackage{jheppub} 

\usepackage[T1]{fontenc} 
\usepackage{graphicx}
\usepackage{epsfig}
\usepackage{subfig}

\title{\boldmath Entangled baryons: violation of Inequalities based on local realism assuming dependence of decays on hidden variables}

\author[b]{Yu Shi,\note{Corresponding author.}}
\author[a]{Ji-Chong Yang}

\affiliation[a]{Department of Physics, Liaoning Normal University, Dalian 116029, China}
\affiliation[b]{Department of Physics, Fudan University, Shanghai, 200433, China}

\emailAdd{yushi@fudan.edu.cn}
\emailAdd{yangjichong@fudan.edu.cn}
\abstract{
Bell inequalities are consequences of local realism while violated by quantum mechanics. In  particle physics, entangled high energy particles can be produced from a common source, and the decay of each particle plays the role of measurement. However, in a hidden variable theory, the decay could be determined by hidden variables. This loophole killed such approaches to Bell test in particle physics. It is a special form of measurement-setting or free-will loophole, which also exists in other systems.  Using  entangled baryons,  we present  new inequalities of local realism with the explicit assumption of the dependence of the decays on hidden variables, as well as the consideration of the statistical mixture of polarizations and the separation of local hidden variables for objects with spacelike distances.  These  violations  closes the measurement-setting loophole once and for all.   We propose to use the processes $\eta _c\to \Lambda \bar{\Lambda}$ and $\chi _{c0} \to \Lambda \bar{\Lambda}$   to test our inequalities, and show that their  violations are likely to be observed with the data already collected in BESIII.}

\begin{document}
\maketitle
\flushbottom

\section{\label{level1}Introduction}

Entanglement in particle physics was noticed long ago~\cite{1960s}, and has since been studied theoretically and  experimentally~\cite{EntangledMesonTheoryAndExp}. The entangled  pseudoscalar mesons are very useful in studying violations of the discrete symmetries~\cite{DoubleTag,CPT}, especially  the time reversal symmetry~\cite{TViolation}. Moreover, many endeavours  have been made to test Bell's inequalities~(BI)~\cite{BI,chsh} using entangled mesons~\cite{MesonBI} and baryons~\cite{BaryonBI,MomemtumRepresentation}.  For these entangled high energy particles,  mostly  the quantum mechanical measurement of each   particle is effectively achieved through its decay, which is not a free choice  of the experimentalist. Therefore, in a realistic or hidden variable theory, the decay could depend on hidden variables at the creation of the entangled  pairs, leading to the violation of BI. In the derivation of BI, however, it is assumed that the decay   does not depend on hidden variables. Therefore,  BI implemented in terms of decays of these entangled high energy  particles cannot serve to distinguish  local realistic theories from quantum mechanics.

Previously it has been noted that an experiment using decay time as the effective measurement basis cannot serve as as a genuine test of BI~\cite{HEPLoopholeDiscussions}. We emphasize that in a realistic theory,  any kind of effective measurement accomplished through  the decay could be determined by the hidden variables.  This is actually a special form of the so-called  measurement-setting for free-will loophole, well known in other systems~\cite{FreeWillLoophole}.

Recently, we made the dependence of the measurement setting on hidden variables an explicit assumption in deriving a new Leggett  inequality~(LI)~\cite{MesonLeggett}, which is a consequence of the so-called crypto-nonlocal realism~\cite{LI,LINature}, and showed   its violation in entangled mesons~\cite{MesonLeggett}. Violation of LI demonstrates that it is not enough to make the realism even cryotp-nonlocal.

In this paper, in terms of entangled hyperons,  a kind of baryons, we present a new kind of inequalities, which are consequences of local realism. But it is different from BI, as it is considered that a  physical state is a statistical mixture of subensembles with definite values of observables, and  that the local hidden variables are separated for objects with spacelike distances, including copies of the same ones from the past when their light cones overlap. In particular, we take into account that  the possibility that the signals, as the effective measurement settings, also depend on hidden variables. Hence our approach closes the measurement setting loopholes once and for all.   Our inequalities are neither LIs, though inspired by them, as we consider local realism, rather than nonlocal realism.

Specifically our inequalities are constructed for  the entangled $\Lambda\bar{\Lambda}$ pairs created in decays of the charmonia $\eta _c$ and $\chi _{c0}$, which are mesons consisting of charm quark $c$ and its antiparticle $\bar{c}$. We estimate the significances of the violations  of our inequities, and find that the violations are  likely to be observed with the data sample collected in BESIII at the Beijing Electron-Position Collider II.

Our proposal demonstrates that the entangled baryon  pairs provide a new playground of entanglement study in the realm of particle physics, for relativistic massive   particles and with  electromagnetic, weak and strong interactions all involved, beyond the scopes of optical and nonrelativistic systems.   As our inequalities are sensitive to the polarization of baryons,  it  can also serve a new  way to study the space-like electromagnetic form factors~(EMFFs) and   polarization effect of hyperons,  which    are   related to the non-zero phase difference~\cite{SpaceLikeEMFF,LambdaPRL,LambdaNature}, and have been studied intensively~\cite{EMFFexpRecently,EMFFthoeryRecently} in order to investigate the charge and magnetization density distributions of a hadron~\cite{ChargeDensityAndEMFF}.

\section{\label{level2} Inequalities for spin-entangled baryons}

\begin{figure}
\includegraphics[scale=0.75]{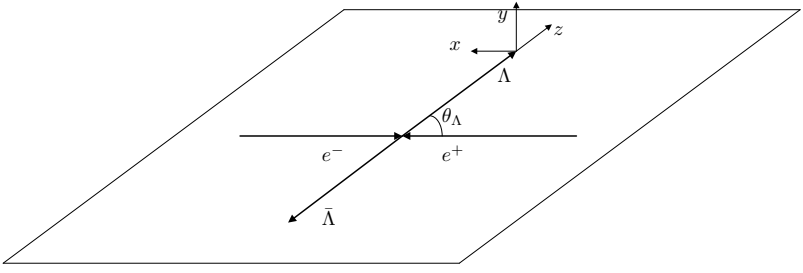}
\caption{\label{fig:coordinate}We first consider the rest frame of the center of mass of the $\Lambda\bar{\Lambda}$ pair,  where   ${\bf z}$ direction is the  direction of the momentum ${\bf p}_{\Lambda}$ of $\Lambda$,   ${\bf y}$ direction is the direction of ${\bf p}_{e^-}\times {\bf p}_{\Lambda}$. By boosting this frame,  the rest frames of $\Lambda$ and $\bar{\Lambda}$  can be obtained repectively.}
\end{figure}

We start with  the angular distributions~\cite{LambdaPRL,LambdaNature}
\begin{equation}
\begin{split}
&\frac{d\sigma (\Lambda \to p \pi ^-)}{d\Omega _p}= \frac{1}{4\pi}\left(1+\alpha _{\Lambda} {\bf s}_{\Lambda}\cdot {\bf n}_p\right),\;\;
\frac{d\sigma(\Lambda \to \bar{p} \pi ^+)}{d\Omega _{\bar{p}}}= \frac{1}{4\pi}\left(1-\alpha _{\Lambda} {\bf s}_{\bar{\Lambda}}\cdot {\bf n}_{\bar{p}}\right),\\
\end{split}
\label{eq.2.1}
\end{equation}
for definite  momentum directions  ${\bf n}_{p}$ (${\bf n}_{\bar{p}}$)   of proton (antiproton)   in the rest frame of  $\Lambda$ ( $\bar{\Lambda}$)  with  definite spin
${\bf s}_{\Lambda}$ ($ {\bf s}_{\bar{\Lambda}}$),      as shown in Fig.~\ref{fig:coordinate},  where $\alpha _{\Lambda}= 0.750 \pm 0.010$ is a constant~\cite{LambdaPRL}, $CP$ violation is ignored.

The angular distribution provides a way to determine
${\bf s}_{\Lambda}$ (${\bf s}_{\bar{\Lambda}}$)  by measuring ${\bf n}_{p}$ (${\bf n}_{\bar{p}}$). Here we use it as a constraint on the hidden variable theories, similar to  Malus' law in defining the polarization vectors existing prior to measurement, valid for photons~\cite{LI} and mesons~\cite{MesonLeggett}.

We consider a  local realistic theory. As Eq.~(\ref{eq.2.1}) implies that the average of  ${\bf n}_p$ equals   $\alpha _{\Lambda} {\bf s}_{\Lambda}/3$ and that of  ${\bf n}_{\bar{p}}$ equals   $-\alpha _{\Lambda} {\bf s}_{\bar{\Lambda}}/3$,   we assume that   in the local realistic theory,  the unit vector signal ${\bf A}$ (${\bf B}$) corresponds to ${\bf n}_p$  (${\bf n}_{\bar{p}}$), and definite polarization vector  ${\bf u}$ (${\bf v}$) corresponds to
${\bf s}_{\Lambda}$ (${\bf s}_{\bar{\Lambda}}$), with
$\bar{\bf A}=\alpha _{\Lambda} {\bf u}/3$ ($\bar {\bf B} = - \alpha _{\Lambda}{\bf v}/3 $),  where the overline denotes the average over all values of the local  hidden variables.

Consider two particles, specifically a  pair of $\Lambda$  and $\bar{\Lambda}$, with spacelike distances.    Indeed,  there are plenty of spacelike events in the $\Lambda\bar{\Lambda}$ experiments.  We   assume that for each of them, the effect of  the polarization   on ${\bf n}_{p}$ (${\bf n}_{\bar{p}}$) is the same as in the  single-particle   case.  Thus for each subensemble with definite polarizations of $\Lambda$ and $\bar{\Lambda}$, we have
\begin{equation}
\begin{split}
&\bar{\bf A}=\int d\lambda _Ad\lambda _B\rho _{A}(\lambda _A)\rho _{B}(\lambda _B){\bf A}(\lambda _A)=\frac{\alpha _{\Lambda}}{3}{\bf u},\\
&\bar{\bf B}=\int d\lambda _Ad\lambda _B\rho _{A}(\lambda _A)\rho _{B}(\lambda _B){\bf B}(\lambda _B)=-\frac{\alpha _{\Lambda}}{3}{\bf v},\\
\end{split}
\label{eq.2.2}
\end{equation}
where  we have separated LHVs to  $\lambda_{A}$  determining $\mathbf{A}$  and $\lambda _{B}$  determining $\mathbf{B}$ with independent  distribution functions  $\rho _A$ and $\rho _B$. In case A and B share some hidden variables from the past when their light cones overlap, in their creation as a pair,   there are copies  of these same hidden variables within  $\lambda_{A}$  and $\lambda _{B}$.

For two arbitrary   unit vectors ${\bf a}$ and ${\bf b}$, we have
\begin{equation}
\begin{split}
&\overline{{\bf A}\cdot {\bf a}{\bf B}\cdot {\bf b}}=\int d\lambda _Ad\lambda _B\rho _{A}(\lambda _A)\rho _{B}(\lambda _B) {\bf A}(\lambda _A)\cdot {\bf a}{\bf B}(\lambda _B)\cdot {\bf b}=-\frac{\alpha ^2_{\Lambda}}{9}{\bf u}\cdot {\bf a}{\bf v}\cdot {\bf b},\\
\end{split}
\label{eq.2.3}
\end{equation}
where Eq.~(\ref{eq.2.2}) has been used.

A physical state is a statistical mixture of subensembles with definite polarization vectors, with  distribution function $F({\bf u},{\bf v})$ in the case of pairs. Thus the correlation function  is
\begin{equation}
\begin{split}
E({\bf a},{\bf b})&\equiv - \langle \overline{{\bf A}\cdot {\bf a}{\bf B}\cdot {\bf b}} \rangle = -\sum_{ij} a_ib_j \langle  \overline{A_i B_j} \rangle \\
& = -\int d{\bf u}d{\bf v} d\lambda _Ad\lambda _B F({\bf u},{\bf v})\rho _{A}(\lambda _A,{\bf u},{\bf v})\rho _{B}(\lambda _B, {\bf u},{\bf v}) {\bf A}_{}(\lambda _A,{\bf u},{\bf v})\cdot {\bf a}{\bf B}_{}(\lambda _B,{\bf u},{\bf v})\cdot {\bf b}\\&=\frac{\alpha ^2_{\Lambda}}{9}\int d{\bf u}d{\bf v} F({\bf u},{\bf v}){\bf u}\cdot {\bf a}{\bf v}\cdot {\bf b},
\end{split},
\label{eq.2.4}
\end{equation}
where a negative sign is used for technical reason,  the dependence of the LHV distributions and signals on  the polarizations are explicitly indicated.

For arbitrary real numbers $-1\leq {\bf u}\cdot {\bf a}\leq 1$ and $-1\leq {\bf v}\cdot {\bf b}\leq 1$, one has ${\bf u}\cdot {\bf a}{\bf v}\cdot {\bf b} \leq 1-|{\bf u}\cdot {\bf a}-{\bf v}\cdot {\bf b}|$,   therefore
\begin{equation}
\begin{split}
&\frac{9}{\alpha ^2_{\Lambda}}E({\bf a},{\bf b})\leq 1-\int d{\bf u}d{\bf v}F({\bf u},{\bf v})|{\bf u}\cdot {\bf a}-{\bf v}\cdot {\bf b}|,\\
\end{split}
\label{eq.2.7}
\end{equation}
the RHS of which first appeared in  a proof of LI~\cite{LI,LINature}. On the plane spanned by ${\bf a}$ and ${\bf b}$, ${\bf a}$ and ${\bf b}$ can be characterized in terms of the azimuth angles as ${\bf a}=(\cos(\phi _a),\sin(\phi _a),0)$ and ${\bf b}=(\cos(\phi _b),\sin(\phi _b),0)$. In terms of $\xi \equiv (\phi _a+\phi _b)/2$ and $\varphi \equiv \phi _b-\phi _a$, the average correlation function  to be measured is  $E^{\bf ab}_N(\varphi)\equiv \sum _{n=1}^N E(2n\pi/N, \varphi)/N$, where $N$ is an integer and $N\geq 2$,   the  superscript ${\bf ab}$   indicates the plane.   This definition  of discrete average avoids the assumption of rotational symmetry~\cite{FairSampling}.  In a way similar to a  proof of LI~\cite{LINature},  we obtain
\begin{equation}
\begin{split}
&\left|E^{\bf ab}_N(\varphi)+E^{\bf ab}_N(0)\right|+\left|E^{\bf cd}_N(\varphi)+E^{\bf cd}_N(0)\right|\leq \frac{\alpha ^2_{\Lambda}}{9}\left(4-2u_N\left|\sin \frac{\varphi}{2}\right|\right),
\end{split}
\label{eq.2.8}
\end{equation}
where $u_N \equiv \cot \left(\pi/2N\right)/N$,    the  superscript ${\bf cd}$ represents a plane   orthogonal to plane ${\bf ab}$.  Note that this inequality for local realistic theories is not based on the dependence of nonlocal variables, as   LI does. Neither is it  BI, as our inequality  additionally assumes polarization vectors and the separation of LHVs, and it combines  various aspects of BI and LI.

We can also obtain an inequality  for the  correlation function defined as  $\hat{E}^{{\bf ab}}_N\equiv \sum _{n=1}^N E(\xi,4n\pi/N)/N$. Writing  ${\bf u}=(\cos(\phi _{u})\sin (\theta _{u}), \sin(\phi _{u})\sin (\theta _{u}), \cos (\theta _{u}))$, and similarly for ${\bf v}$, we   rewrite Eq.~(\ref{eq.2.7}) as
\begin{equation}
\begin{split}
&\frac{9}{\alpha ^2_{\Lambda}}E(\xi,\varphi)\leq 1- 2\int _0^{2\pi}\sin \theta _u d\theta _u\int _0^{2\pi} d\psi\int _0^{\pi}\sin \theta _v d\theta _v \int _0^{2\pi}d\chi F(\theta _u, \theta_v, \chi, \psi)\\
&\times \left|a_2\cos\frac{\varphi - \chi}{2}\cos(\xi-\psi)-a_1\sin\frac{\varphi - \chi}{2}\sin(\xi -\psi)\right|,
\end{split}
\label{eq.2.9}
\end{equation}
in a way similar to  Eq.~(27) in the  supplement of Ref.~\cite{LINature}.
With $a_1 \equiv (\sin \theta _u +\sin \theta _v)/2$, $a_2 \equiv(\sin \theta _u -\sin \theta _v)/2$, $\psi \equiv (\phi _u+\phi _v)/2$ and $\chi \equiv\phi _u-\phi _v$, we have $\left|a_2\cos\frac{\varphi - \chi}{2}\cos(\xi-\psi)-a_1\sin\frac{\varphi - \chi}{2}\sin(\xi -\psi)\right|\\
=\sqrt{a_2^2\cos^2(\xi-\psi)+a_1^2\sin^2(\xi-\psi)} \left|\cos \left(\frac{\varphi -\chi}{2}+\alpha\right)\right|$,
where $\alpha$ is some constant real number. Consequently
\begin{equation}
\begin{split}
&\frac{9}{\alpha ^2_{\Lambda}}\frac{1}{N}\sum _{n=1}^N E(\xi,\frac{4n\pi}{N})\leq 1- 2u_N\int _0^{2\pi}\sin \theta _u d\theta _u\int _0^{2\pi} d\psi\int _0^{\pi}\sin \theta _v d\theta _v \int _0^{2\pi}d\chi F(\theta _u, \theta_v, \chi, \psi)\\
&\times \sqrt{a_2^2\cos^2(\xi-\psi)+a_1^2\sin^2(\xi-\psi)}.
\end{split}
\label{eq.2.11}
\end{equation}
Then following the method in  Ref.~\cite{LINature},  we obtain
\begin{equation}
\begin{split}
&\left|\hat{E}^{\bf ab}_N(\xi)+\hat{E}^{\bf ab}_N(0)\right|+\left|\hat{E}^{\bf cd}_N(\xi)+\hat{E}^{\bf cd}_N(0)\right|\leq \frac{\alpha ^2_{\Lambda}}{9}\left(4-2u_N\left|\sin \xi\right|\right),
\end{split}
\label{eq.2.12}
\end{equation}
where the superscripts ${\bf ab}$ and ${\bf cd}$ indicate   orthogonal planes.

Note that in the local realistic theory leading to our inequalities, the state of two particles are generically a statistical  mixture of subensembles with definite polarizations. The case with definite polarizations is only a special case. In contrast, a
previous BI for $\Lambda$ and $\bar{\Lambda}$ was based on the  assumption of  definite polarizations~\cite{MomemtumRepresentation}.

\section{\label{level3} Violations  of our inequalities }

Now we show that the above  two inequalities are violated by quantum mechanics and the standard model of particle physics.  For simplicity, we set  $N=4$. The significance of the violation is estimated by using  a violation  ratio defined as $r \equiv \left(\left|L_{QM}\right|-\left|R\right|\right)/\left|L_{QM}\right|$, where $L_{QM}$ is the quantum mechanical result of  the LHS of the inequality, $R$ represents the RHS of the inequality.   For example, for the first inequality    Eq.~(\ref{eq.2.8}), $R=\alpha ^2_{\Lambda}\left(4-2u_N\left|\sin \left(\varphi / 2\right)\right|\right)/9$, and if we choose ${\bf a}{\bf b}$ on to be  ${\bf xy}$ plane and ${\bf c}{\bf d}$ to be the ${\bf xz}$ plane, then  $L_{QM}=\left|E^{\bf xy}_4(\varphi)+E^{\bf xy}_4(0)\right|+\left|E^{\bf xz}_4(\varphi)+E^{\bf xz}_4(0)\right|$. Obviously $r\leq 0$ means that the inequality is satisfied.

\subsection{\label{level3.1}The process with $\eta _c$ and $\chi _{c0}$}

\begin{figure}
\includegraphics[scale=0.75]{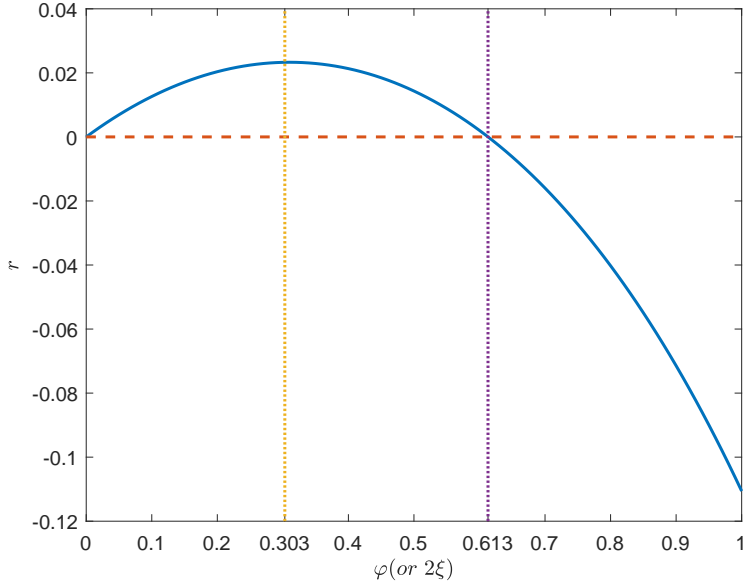}
\caption{\label{fig:etac} The violation ratio $r$ for the first (second) inequality,  as a function of $\varphi$ ($2\xi$ ) for $\eta _c$ ($\chi _{c_0}$).  }
\end{figure}

Consider  $\eta _c$ and $\chi _{c0}$ processes, where  $\eta _c$ and $\chi _{c0}$ are spinless. They are indicated as   superscripts in various quantities below. Using the decay amplitude given in Ref.~\cite{MomemtumRepresentation}, $\mathcal{M}_{\eta _c}=\mathcal{M}_{\Lambda}\bar{u}(p_{\Lambda},s_{\Lambda})
\gamma _5 v(p_{\bar{\Lambda}},s_{\bar{\Lambda}})
\mathcal{M}_{\bar{\Lambda}}$, $\mathcal{M}_{\chi _{c0}}=\mathcal{M}_{\Lambda}
\bar{u}(p_{\Lambda},s_{\Lambda})
v(p_{\bar{\Lambda}},s_{\bar{\Lambda}})
\mathcal{M}_{\bar{\Lambda}}$,  $\mathcal{M}_{\Lambda}=
\bar{u}(p_p,s_p)\left(1+c_{\Lambda}\gamma _5\right)u(p_{\Lambda},s_{\Lambda})$, $
\mathcal{M}_{\bar{\Lambda}}=
\bar{v}(p_{\bar{\Lambda}},s_{\bar{\Lambda}})
\left(1-c_{\bar{\Lambda}}\gamma_5\right)
v(p_{\bar{p}},s_{\bar{p}})$, where the notations are standard, we find the joint angular distributions
\begin{equation}
\begin{split}
&\frac{d\sigma ^{\eta _c}}{d\Omega _p\Omega _{\bar{p}}}\propto 1+\alpha_{\Lambda} ^2{\bf n}_p\cdot {\bf n}_{\bar{p}},\;\;\;\frac{d\sigma ^{\chi _{c0}}}{d\Omega _p\Omega _{\bar{p}}}\propto 1-\alpha_{\Lambda} ^2\left(n_{px}n_{\bar{p}x}+
n_{py}n_{\bar{p}y}-n_{pz}n_{\bar{p}z}\right).\\
\end{split}
\label{eq.3.2}
\end{equation}
Then we find that  for $\eta _c$ processes, the correlation function  $E^{\eta _c}({\bf a},{\bf b})$ is independent of the plane we choose, while for  $\chi _{c0}$ processes, we can choose the ${\bf xz}$ and ${\bf yz}$ planes  such that the  correlation functions are of a same form,
\begin{equation}
\begin{split}
&E^{\eta _c}(\xi,\varphi)=E^{\eta _c}_4(\varphi)=-\frac{\alpha ^2_{\Lambda}}{9}\cos (\varphi),\;\;E^{\chi _{c0}}(\xi,\varphi)=\hat{E}_4^{\chi _{c0}}(\xi)=\frac{\alpha ^2_{\Lambda}}{9}\cos (2\xi).\\
\end{split}
\label{eq.3.3}
\end{equation}
Consider the  $\eta _c$ process. The first inequality  Eq.~(\ref{eq.2.8}) implies $L^{\eta _c}=2\alpha ^2_{\Lambda}\left|\cos(\varphi) + 1\right|/9$, thus  the maximum of the violation ratio is $r_m=u_N^2/\left(16-u_N^2\right)\approx 0.0233$,   at $\varphi _m = 2 \tan^{-1}\left(u_N/\sqrt{16-u_N^2}\right)\approx 0.303$, as depicted in Fig.~\ref{fig:etac}.  Similarly, consider the $\chi_{c0}$ process. For the violation of the second inequality   Eq.~(\ref{eq.2.12}),    the maximal violation ratio   $r_m$ is same as $\eta _c$,   at $\xi_m = \varphi _m / 2$. We also note that the first inequality cannot be violated in the  $\chi_{c0}$  process while the second inequality cannot be violated in the $\eta _c$  process.

\subsection{\label{level3.2}The process with polarization effects }

Now we consider the process $e^+e^-\to \Lambda \bar{\Lambda}\to p\pi ^0 \bar{p}\pi ^+$ and $e^+e^-\to J/\Psi \to \Lambda \bar{\Lambda}\to p\pi ^0 \bar{p}\pi ^+$, with polarizations. The joint angular distribution can be parameterized as~\cite{LambdaPRL,LambdaNature}
\begin{equation}
\begin{split}
&\frac{d\sigma}{d\Omega _{\Lambda}d\Omega _pd\Omega _{\bar{p}}}\propto 1+\eta \cos^2 \theta _{\Lambda} - \alpha^2 _{\Lambda}\left(\sin ^2\theta _{\Lambda}\left(n_{px}n_{\bar{p}x}-\eta n_{py}n_{\bar{p}y}\right)+\left(\cos ^2\theta _{\Lambda}+\eta\right)n_{pz}n_{\bar{p}z}\right)\\
&-\alpha^2 _{\Lambda}\sqrt{1-\eta ^2}\cos (\Delta \Phi)\sin \theta _{\Lambda}\cos \theta _{\Lambda}\left(n_{px}n_{\bar{p}z}+n_{pz}n_{\bar{p}x}\right)\\
&+\alpha _{\Lambda}\sqrt{1-\eta ^2}\sin (\Delta \Phi)\sin \theta _{\Lambda}\cos \theta _{\Lambda} (n_{py}-n_{\bar{p}y}),\\
\end{split}
\label{eq.3.4}
\end{equation}
where $n_{px}$ is the x-component of $n_p$, and so on,   $\theta _{\Lambda}$ is the angle between momenta of $\Lambda$ and  $e^-$,  as shown in Fig.~\ref{fig:coordinate}, $\eta$ and $\Delta \Phi$ are parameters related to polarization effects. It has been noticed that, the maximal violation of BI is related to degree of entanglement~\cite{ViolationOfBI}. We find that the violation of BI given in  Ref.~\cite{MomemtumRepresentation} reaches the maximum when $\theta _{\Lambda} = \pi / 2$, where the polarization effect is minimal. Therefore we consider this region, where incidently the event number is found to be  large in experiments~\cite{LambdaPRL,LambdaNature}. Hence we  only consider these events, for which
\begin{equation}
\begin{split}
&\frac{d\sigma (\theta _{\Lambda}=\frac{\pi}{2})}{d\Omega _pd\Omega _{\bar{p}}}\propto 1 - \alpha^2 _{\Lambda}\left(n_{px}n_{\bar{p}x}-\eta n_{py}n_{\bar{p}y}+\eta n_{pz}n_{\bar{p}z}\right)\\
\end{split}
\label{eq.3.5}
\end{equation}
Therefore we find
\begin{equation}
\begin{split}
&\hat{E}^{\bf xy}_4(\xi)=-\frac{\alpha ^2_{\Lambda}}{9}\eta \cos (2\xi),\;\;\hat{E}^{\bf zy}_4(\xi)=-\frac{\alpha ^2_{\Lambda}}{9}\frac{1+\eta}{2} \cos (2\xi).
\end{split}
\label{eq.3.6}
\end{equation}

\begin{figure}
\includegraphics[scale=0.75]{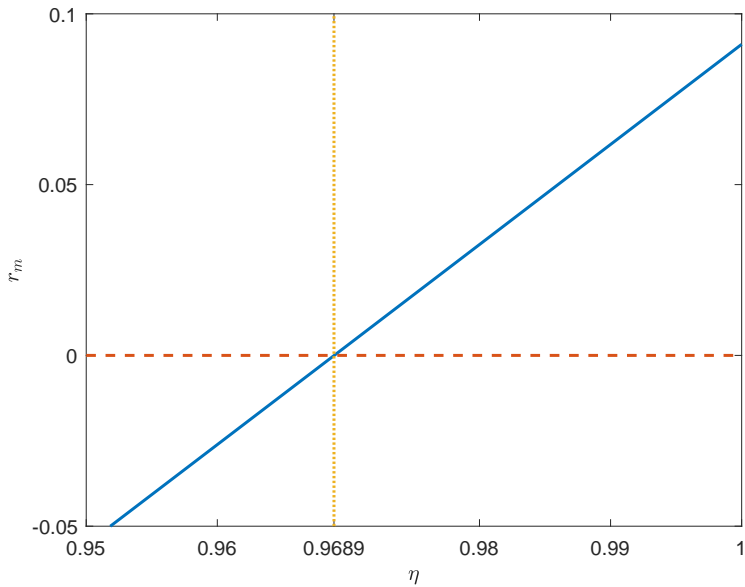}
\caption{\label{fig:jpsi}The maximum $r_m$ of the violation ratio $r$,  as a function of the parameter $\eta$,  for entangled $\Lambda\bar{\Lambda}$ pairs with polarization effects.}
\end{figure}

If our second inequality  Eq.~(\ref{eq.2.12}) is  violated, the   violation is maximal at $\xi = \pi - \tan ^{-1}\left(u_N/\sqrt{\left(1+3\eta\right)^2-u_N^2}\right)$. For this maximal violation ratio to be positive, the necessary condition is  $\eta > \left(1+\sqrt{4-u_N^2}\right) / 3 \approx 0.97$, as shown in Fig.~\ref{fig:jpsi}.  However, it is known  from the experiments that $\eta = 0.46$ for $e^+e^-\to J/\psi \to \Lambda \bar{\Lambda}$~\cite{LambdaNature},  and $\eta = 0.12$  for $e^+e^-\to \Lambda \bar{\Lambda}$~\cite{LambdaPRL}. Therefore this inequality cannot be violated in either case. Besides, for any $\eta$,   the first inequality   Eq.~(\ref{eq.2.8}) cannot be violated.

\section{\label{level5}Summary and  discussions}

In this Letter, we consider local realistic theories with the specifications that  the local  hidden variables for different objects with spacelike distances are separated and that the physical states are statistical mixtures of subensembles with definite polarizations. We present two inequalities that are shown to be violated by entangled baryons.

In the usual BI test using entangled spins or polarizations, one needs to choose the guide   axis of the  measurement.  When the choices of the two guild axes are not independent, or only limited choices of the axes are allowed, or the guide axes are determined by hidden variables, it is possible that even a local realistic  theory can violate BI.  This measurement-setting or free-will loophole has been a general defect in most of the previous approaches to BI based on decays of high energy particles. In using $\Lambda \to p\pi ^-$ ($\bar{\Lambda} \to \bar{p}\pi ^+$), where the momentum direction of the proton (antiproton) acts as an effective guide  axis for the  the spin of $\Lambda$  ( $\bar{\Lambda}$),  the momentum of the proton (antiproton) cannot be freely set by the experimentalists, and could be determined by  hidden variables carried over from the generation of the entangled particle.  All these possibilities are different manifestations of  measurement or free-will setting loophole.

In the local realistic  theories considered here, the  dependence of the guide axes, or  the  momenta of the protons and  the antiprotons, on the hidden variables is taken as an assumption in deriving the inequalities.   Therefore the violations of these inequalities close the measurement-setting or free-will loophole  once and for all.

We find that for $\eta _c\to \Lambda \bar{\Lambda}$ and $\chi _{c0}\to \Lambda \bar{\Lambda}$, our inequalities can be violated. For  $e^+e^-\to \Lambda \bar{\Lambda}$ and $e^+e^-\to J/\Psi\to \Lambda \bar{\Lambda}$, the inequalities are sensitive to the polarization effect, and cannot be violated.

We propose to   test our inequalities in  experiments. The relative significance  of the violation of the first inequality is $r_m\approx 0.0233$. Typically, to observe a relative significance  at the order of $10^{-2}$, the number of events are required to be at the order of $1/r_m^2 \sim 10^4$. For example, the $\eta _c$ can be produced from $J/\Psi\to \gamma \eta _c$ at BESIII, with the branch  ratio   ${\rm Br}\left(J/\Psi\to \gamma\left( \eta _c\to \Lambda \bar{\Lambda}\to p\pi^- \bar{p} \pi^+\right)\right)=9.8\pm 2.6\times 10^{-6}$~\cite{MomemtumRepresentation}. A data sample of  $10$ billion  $J/\Psi$ events has been collected by BESIII~\cite{li}, updating the $1.31\times 10^9$ events used in the previous analysis~\cite{LambdaNature}. $\eta_c$ and $\chi_c$ processes, with event numbers up to millions and tens of thousands respectively,  are also under analyses in BESIII~\cite{li}. It is likely that the violation of our inequalities can be tested using these  data.

\begin{acknowledgments}
We thank Prof. Haibo Li from BESIII collaboration for useful discussion. This work is supported by National Science Foundation of China (Grant No. 11574054).
\end{acknowledgments}

\end{document}